\documentclass[aps,twocolumn,preprintnumbers,nofootinbib,superscriptaddress]{revtex4-1}
\usepackage{amsmath} \usepackage{graphicx} \usepackage{amsfonts}
\usepackage{array} \usepackage{amsthm} \usepackage{bm}
\usepackage{palatino} \usepackage{mathpazo} 
\usepackage{supertabular}

\usepackage[breaklinks]{hyperref}
\usepackage{color}

\newcommand{\nn}{\nonumber}
\newcommand{\be}{\begin{equation}}
\newcommand{\ee}{\end{equation}}
\newcommand{\ba}{\begin{eqnarray}}
\newcommand{\ea}{\end{eqnarray}}
\newcommand{\bal}{\begin{align}}
\newcommand{\eal}{\end{align}}

\newcommand{\e}{\text{e}}
\newcommand{\dd}{\text{d}}
\newcommand{\ii}{\text{i}}
\newcommand{\bb}{\bibitem}

\newcommand{\bw}{\begin{widetext}}
\newcommand{\ew}{\end{widetext}}

\begin{document}
\title{Dynamical and static solutions to $R=0$-scalar-tensor theory}

\author{Mustapha Azreg-A\"{\i}nou}
\affiliation{Ba\c{s}kent University, Engineering Faculty, Ba\u{g}l{\i}ca Campus, Ankara, Turkey}


\begin{abstract}
We consider the most cosmologically interesting and relevant case of scalar-tensor theory (STT) and derive new normal and phantom, dynamical and static, solutions. We determine the Bianchi I Kasner exponents and show that the dynamical solutions are heteroclinic orbits connecting two singularities. Approaching the singularities, a purely transverse expansion (no radial expansion or collapse) may occur.
\end{abstract}


\maketitle

\section{The \pmb{$R=0$}-scalar-tensor theory}
Let us start with the class of scalar-tensor theory (STT) action describing the dynamics of a conformally coupled massless scalar field $\sigma$
\begin{equation}\label{1}
S=\int \dd ^4x\sqrt{|g|}\Big(\phi(\sigma) R-\frac{1}{2}g^{\mu\nu}\sigma_{,\mu}\sigma_{,\nu}\Big),
\end{equation}
where $\phi(\sigma)$ is a conformal factor function of $\sigma$ and $R^{\mu}_{\ \ \nu\delta\eta}=\Gamma^{\mu}_{\nu\eta,\delta}-\Gamma^{\mu}_{\nu\delta,\eta}+\cdots$ ($\mu:\,1\to 4$). This action is equivalent to the Brans--Dicke-like action
\begin{equation}\label{2}
S=\int \dd ^4x\sqrt{|g|}\Big(\phi R-\frac{\omega (\phi)}{\phi}g^{\mu\nu}\phi_{,\mu}\phi_{,\nu}\Big),
\end{equation}
with
\begin{equation*}
\omega (\phi)=\frac{1}{2}~\frac{\phi}{(\phi_{,\sigma})^2},
\end{equation*}
where $\phi_{,\sigma}$ is to be expressed in terms of $\phi$ upon reversing the function $\phi(\sigma)$.

The equations of motion derived from~\eqref{1} take the form
\begin{multline}\label{3}
\phi \Big(R_{\mu\nu}-\frac{1}{2}Rg_{\mu\nu}\Big)+g_{\mu\nu}\Big(\phi_{,\sigma\sigma}+\frac{1}{4}\Big)\sigma_{,\delta}\sigma^{,\delta}\\-\Big(\phi_{,\sigma\sigma}+\frac{1}{2}\Big)\sigma_{,\mu}\sigma_{,\nu}
+\phi_{,\sigma}(g_{\mu\nu}\square\sigma-\nabla_{\nu}\nabla_{\mu}\sigma)=0,
\end{multline}
\begin{equation}\label{4}
\square\sigma +\phi_{,\sigma}R=0.
\end{equation}
The case with vanishing scalar curvature $R=0$ has received much attention in cosmological studies~\cite{trans,duality}. Assuming $R=0$, Eq.~\eqref{4} yields
\begin{equation}\label{5}
\square\sigma =0,
\end{equation}
and \eqref{3} reduces to
\begin{multline}\label{6}
H_{\mu\nu}\equiv\phi R_{\mu\nu}+g_{\mu\nu}\Big(\phi_{,\sigma\sigma}+\frac{1}{4}\Big)\sigma_{,\delta}\sigma^{,\delta}\\-\Big(\phi_{,\sigma\sigma}+\frac{1}{2}\Big)\sigma_{,\mu}\sigma_{,\nu}
-\phi_{,\sigma}\nabla_{\nu}\nabla_{\mu}\sigma=0,
\end{multline}
the trace of which implies
\begin{equation}
(6\phi_{,\sigma\sigma}+1)\sigma_{,\delta}\sigma^{,\delta}=0.
\end{equation}
The case $\sigma=\text{const.}$ corresponds to Einstein's general relativity and we drop this case from considerations. Thus, $R=0$ corresponds to $\phi(\sigma)=\phi_0-(\sigma-\sigma_0)^2/12$ where ($\phi_0,\,\sigma_0$) are real constants. We can drop $\sigma_0$ by redefining $\sigma$ in~\eqref{1}: $\sigma\to\sigma-\sigma_0$ and this amounts to drop the linear term in the expression of $\phi(\sigma)$
\begin{equation}\label{7}
\phi(\sigma)=\phi_0-\frac{\sigma^2}{12}\qquad\text{and}\qquad \phi_0\in\mathbb{R}.
\end{equation}
The dimensionless Dicke coupling function corresponding to $R=0$ reads
\begin{equation}\label{omega}
\omega(\phi)=\frac{3}{2}~\frac{\phi}{\phi_0-\phi}.
\end{equation}

We need to set further conditions on the sign of $\phi$. First of all we let $\epsilon\equiv \text{sign}(\phi_0)$ and $V_0\equiv |\phi_0|$ so that $\phi_0=\epsilon V_0$. The case $\phi>0$,
\begin{equation}\label{15a}
\epsilon =+1\quad\text{and}\quad -\sqrt{12V_0}<\sigma<\sqrt{12V_0},
\end{equation}
corresponds to normal solutions while the case $\phi<0$
\begin{equation}\label{15b}
\epsilon =-1\quad\text{and}\quad -\infty<\sigma<\infty,
\end{equation}
corresponds to phantom solutions.

On setting
\begin{equation}\label{x}
x\equiv \sigma/\sqrt{12 V_0},
\end{equation}
and using~\eqref{7} we bring~\eqref{6} and~\eqref{5} to the forms
\begin{align}\label{6b}
&(\epsilon -x^2) R_{\mu\nu}+g_{\mu\nu}x_{,\delta}x^{,\delta}-4x_{,\mu}x_{,\nu}
+2x\nabla_{\nu}\nabla_{\mu}x=0,\nn\\
&\square x =0.
\end{align}
Now, consider the conformal transformation
\begin{equation}\label{6c}
\tilde{g}_{\mu\nu}=\Omega^2g_{\mu\nu}=x^2g_{\mu\nu},\qquad \tilde{x}=x^{-1}.
\end{equation}
Using the laws of transformation of $R_{\mu\nu}$ and $\square$, given by
\begin{align}
&\tilde{R}_{\mu\nu}=R_{\mu\nu}+\Omega^{-2}(4\Omega_{,\mu}\Omega_{,\nu}-g_{\mu\nu}\Omega_{,\delta}\Omega^{,\delta})\nn\\
&\qquad -\Omega^{-1}(2\nabla_{\nu}\nabla_{\mu}\Omega+g_{\mu\nu}\square\Omega),\\
&\tilde{\square} (\cdot) =\Omega^{-2}[\square (\cdot)+2\Omega^{-1}g^{\mu\nu}\Omega_{,\mu}(\cdot)_{,\nu}],
\end{align}
it is straightforward to show that the equations~\eqref{6b} remain invariant under the conformal transformation~\eqref{6c}, that is,
\begin{align}\label{6d}
&(\epsilon -\tilde{x}^2) \tilde{R}_{\mu\nu}+\tilde{g}_{\mu\nu}\tilde{x}_{,\delta}\tilde{x}^{,\delta}-4\tilde{x}_{,\mu}\tilde{x}_{,\nu}
+2\tilde{x}\tilde{\nabla}_{\nu}\tilde{\nabla}_{\mu}\tilde{x}=0,\nn\\
&\tilde{\square} \tilde{x} =0.
\end{align}
This implies that if ($g_{\mu\nu},\,x$) is a solution to~\eqref{6b} then ($x^2g_{\mu\nu},\,x^{-1}$) is another solution to~\eqref{6b}~\cite{conf,conf2}.

To pursue the integration of~(\ref{5}, \ref{6}) or ~\eqref{6b} we introduce the spherically symmetric ansatz ($\dd \Omega^2\equiv\dd\theta^2+\sin ^2\theta\dd\varphi^2$),
\begin{equation}\label{8}
\dd s^2_{\text{J}}=\e ^{2\gamma(u)}\dd t^2 - \e ^{2\alpha(u)}\dd u^2 - \e ^{2\beta(u)}\dd \Omega^2
\end{equation}
and $\sigma\equiv\sigma(u)$, in the Jordan frame. Since the shape of the function $\alpha(u)$ can be changed by a coordinate transformation involving only the radial coordinate $u$, we have the freedom to choose $\alpha(u)$ to obey the harmonic gauge,
\begin{equation}\label{9}
\alpha = \gamma + 2\beta .
\end{equation}
The purpose of working with this gauge is that the Laplace--Beltrami operator reduces to $\square(\cdot)=-\e ^{-2(\gamma + 2\beta)}(\cdot)_{,uu}$, so that~\eqref{5} yields $\sigma=Cu+C_1$ where ($C,\,C_1$) are real constants. Assuming that $\sigma=0$ and $u=0$ correspond to spatial infinity, we are led to
\begin{equation}\label{10}
\sigma = C u\qquad\text{and}\qquad C\in\mathbb{R}.
\end{equation}
The non-vanishing field equations~\eqref{6} take the form:
\begin{align*}
&H_{11}=\frac{{\rm e}^{-4 \beta }}{12} \big[12 {\rm e}^{2 (\beta +\gamma )} \phi +12 \phi (\beta '^2+2 \beta
	' \gamma '-2 \beta '')\\
&-C (C+2 \sigma  \gamma ')\big]=0,\\
&H_{22}=-\frac{C \sigma }{6} (2 \beta '+\gamma ')+\phi  \big(-{\rm e}^{2 (\beta +\gamma )}+\beta '^{\,2}+2 \beta ' \gamma '\big)\\
&-\frac{C^2}{4}=0,\\
&H_{33}=\frac{H_{44}}{\sin ^2\theta }=\frac{{\rm e}^{-2 (\beta +\gamma )}}{12}  \big[2 \beta ' (C \sigma -12 \phi  \gamma
	')-12 \phi  \beta '^{\,2}\\
&+12 \phi  (\beta ''+\gamma '')+C^2\big]=0,
\end{align*}
where the prime indicates taking derivative with respect to $u$.

As we mentioned earlier, the case $\sigma=\text{const.}$ corresponds to Einstein's general relativity and it yields the Schwarzschild BH. So, from now on, we assume $C\neq 0$. In this case one can use $\sigma$ as a radial coordinate. Using $\sigma$ as a radial coordinate, the field equations~\eqref{6} and the constraint $R=0$ take the following forms.
\begin{equation}
\frac{\e ^{4\beta}H_{11}+H_{22}}{2}+\e ^{2(\beta+\gamma)}H_{33}=0\Rightarrow (12\phi\gamma_{,\sigma})_{,\sigma}=1.
\end{equation}
This first yields
\begin{equation}\label{11}
\gamma_{,\sigma}=\frac{\sigma+C_2}{12\epsilon V_0-\sigma^2}\qquad\text{and}\qquad C_2\in\mathbb{R},
\end{equation}
and it suggests that
\begin{equation}\label{12}
z(\sigma)\equiv 12\phi\beta_{,\sigma}
\end{equation}
might be a suitable variable for the integration of~\eqref{6} instead of $\beta$, which is expressed in terms of $z$ by
\begin{equation}\label{13}
\e ^{4\beta}H_{11}-H_{22}=0\Rightarrow \e^{2 \beta }=\frac{C^2(z_{,\sigma }-1)}{12 \phi }~\e^{-2 \gamma }.
\end{equation}
With the aid of~\eqref{11}, \eqref{12} and $\beta_{,\sigma\sigma}=(\phi z_{,\sigma}-z \phi_{,\sigma})/12\phi^2$, we find that $z(\sigma)$ satisfies the inhomogeneous Bernoulli differential equation
\begin{multline}\label{14}
\e ^{4\beta}H_{11}+H_{22}=0\Rightarrow\\
(12 \epsilon V_0-\sigma ^2) z_{,\sigma }+2 (\sigma -C_2) z=z^2-2 C_2 \sigma -24 \epsilon V_0.
\end{multline}
On setting
\begin{equation}\label{def}
y=\frac{z}{\sqrt{12 V_0}}\quad \text{and}\quad  \eta = \frac{C_2}{\sqrt{12 V_0}},
\end{equation}
along with~\eqref{x} we bring the differential equation~\eqref{14} to the much simpler expression
\begin{equation}\label{14b}
(\epsilon -x^2) y_{,x}+2 (x -\eta) y=y^2-2 \eta x -2\epsilon .
\end{equation}

The remaining steps consist in solving~\eqref{11} to determine $\gamma(\sigma)$ and~\eqref{14}, or~\eqref{14b}, to determine $z(\sigma)$. Finally we evaluate $\e^{2 \beta }$ from~\eqref{13} using the expressions of ($\gamma,\;z$) and~\eqref{7}.

We can now determine the value of $C$~\eqref{10} using the constraint $R=0$. This constraint upon expressing the term $\e^{2(\beta+\gamma)}$ using~\eqref{13} reduces to
\begin{equation*}
	z_{,\sigma }-1+12 \phi  C^2 (\beta _{,\sigma }^2+2 \beta _{,\sigma } \gamma _{,\sigma }-2 \beta _{,\sigma \sigma }-\gamma _{,\sigma \sigma
	})=0.
\end{equation*}
Now, with the aid of~\eqref{11}, \eqref{12} and $\beta_{,\sigma\sigma}=(\phi z_{,\sigma}-z \phi_{,\sigma})/12\phi^2$ we can first eliminate all the derivatives of ($\beta,\,\gamma$), then using the differential equation~\eqref{14} to eliminate $z_{,\sigma }$ we arrive at
\begin{equation}
	\hspace{-1.0mm}(C^2-1) [z^2+2 (C_2-\sigma ) z+\sigma  (\sigma -2 C_2)-36 \epsilon V_0]=0.
\end{equation}
Since the algebraic expression inside the square parentheses does vanish only for \emph{linear} functions $z(\sigma)$:
\begin{equation}
z=\sigma -C_2\pm \sqrt{C_2^2+36\epsilon V_0},
\end{equation}
which, by~\eqref{13}, would yield an identically zero metric component $\e ^{2\beta}$. The whole metric is undefined in this case which we exclude from our analysis. We thus conclude that
\begin{equation}\label{key}
	C=\pm 1.
\end{equation}
Without loss of generality, we opt for $C=1$ and thus $\sigma=u$.

Once the final solution $\dd s^2_{\text{J}}$ in the Jordan frame is known, we obtain the solution in the Einstein frame $\dd s^2_{\text{E}}$ via the conformal transformation
\begin{equation}\label{trans}
\dd s^2_{\text{J}}= (\phi_0/\phi)\dd s^2_{\text{E}}.
\end{equation}
In this work we consider the Einstein frame to be the physical one relying on the argument given in~\cite{arg1,arg2,arg3}. In the linearized version of the theory $g_{\text{J}\ \mu\nu}\simeq \eta_{\mu\nu}+h_{\mu\nu}$ and $g_{\text{E}\ \mu\nu}\simeq \eta_{\mu\nu}+\rho_{\mu\nu}$ where $\eta_{\mu\nu}$ is the Minkowski metric. The argument advanced in~\cite{arg1,arg2,arg3} is that the spin 2 gravitational field is described by the Einstein frame corrections $\rho_{\mu\nu}$ to the flat metric while $h_{\mu\nu}$ describes a mixture of spin 0 and spin 2 fields.

For the remaining sections of this work we assume
\begin{equation}\label{ass}
\eta \geq 0.
\end{equation} 

\section{Normal solutions}
We assume that~\eqref{15a} holds. Eqs.~(\ref{11}, \ref{14b}) are solved by
\begin{align}
&\label{16a}\e^{2 \gamma }=\frac{1}{12 \phi } \Big(\frac{1+x}{1-x}\Big)^{\eta },\\
&\label{16b}y=x -\eta -k_+\tanh \Big(k_+\,\text{arctanh}\,x-C_3\,\ii\Big),
\end{align}
where we have set
\begin{equation}
k_+\equiv \sqrt{\eta^2+3}\geq \sqrt{3}.
\end{equation}
An additive constant in the expression of $\gamma$ has been set to zero by re-parameterizing the time coordinate. The new constant $C_3$ is complex and $\ii^2=-1$.

The r.h.s of the second expression in~\eqref{13} is positive for static solutions and negative for dynamical solutions. All that depends on the value of $C_3$.

\subsection{\textbf{Static normal solutions: $\pmb{C_3=(2\ell+1)\pi/2+\lambda\,\ii}$, $\pmb{\ell\in\mathbb{Z}}$, $\pmb{\lambda\in\mathbb{R}}$.}} For this value of $C_3$ the final expressions of ($y,\,\e^{2\beta},\,\e^{2\alpha}=\e^{4\beta}\e^{2\gamma}$) are
\begin{align}
&\label{17a}y=x -\eta -k_+\tanh \Big(k_+\,\text{arctanh}\,x\Big),\\
&\label{17b}\e^{2\beta}=\frac{V_0}{12\phi^2}~k_+^2\text{csch}^2\Big(k_+\,\text{arctanh}\,x\Big)~\e^{-2\gamma},\\
&\label{17c}\e^{2\alpha}=\frac{V_0^2}{12^2\phi^4}~k_+^4\text{csch}^4\Big(k_+\,\text{arctanh}\,x\Big)~\e^{-2\gamma},
\end{align}
where we have set $\lambda=0$, as this is always possible, $\text{csch}(\cdot)=1/\sinh(\cdot)$, and $\e^{2\gamma}$ is given by~\eqref{16a}. The static metric is given by~\eqref{8} where $t$ is timelike and the other coordinates are spacelike.

From~\eqref{16a} and~\eqref{17b} we see that if $\e^{2\gamma}$ has a zero, that is an event horizon, $\e^{2\beta}$ and the horizon area diverge implying a vanishing Hawking temperature.

To bring the metric~\eqref{16a}, \eqref{17b} and~\eqref{17c} to more familiar form we define the new radial coordinates $\rho$ and $r$ as follows. Defining $\rho$ by
\begin{equation}
\e^{-2\rho} = \frac{\sqrt{12 V_0}+\sigma }{\sqrt{12 V_0}-\sigma }= \frac{1+x}{1-x},
\end{equation}
yields
\begin{align}\label{metric}
&\dd s^2_{\text{J}}=\cosh^2\rho~\dd s^2_{\text{E}},\;\phi =\frac{\phi_0}{\cosh^2\rho},\; \frac{\sigma}{\sqrt{12\phi_0}}=-\tanh \rho ,\nonumber\\
&\dd s^2_{\text{E}}=\e ^{-2\eta\rho}\dd t^2 - \frac{k_+^2\e ^{2\eta\rho}}{\sinh^2(k_+\rho)}\Big(\frac{k_+^2\dd \rho^2}{\sinh^2(k_+\rho)} +\dd \Omega^2\Big),
\end{align}
where we have made the substitution $t\to \sqrt{12\phi_0}\,t$. The metric $\dd s^2_{\text{E}}$ was derived in Refs~\cite{cold1,cold2} in the special case $\omega=\text{const.}$, while for the solution~\eqref{metric} $\omega$ is not constant and is given by~\eqref{omega}. The general solution was later derived in~\cite{stt}.

Following~\cite{cold1,cold2} we define the radial coordinate $r$ by
\begin{equation}\label{r}
\e ^{-2k_+\rho}=f(r)=1-\frac{2k_+}{r},
\end{equation}
the metric~\eqref{metric} and the corresponding potential $\sigma$ along with the metric and potential obtained by the conformal transformation~\eqref{6c} take the forms
\begin{align}\label{metric2}
&\dd s^2_{\text{J}}=\frac{(1\pm f^{1/k_+})^2f^{-1/k_+}}{4}~\dd s^2_{\text{E}},\nn\\
&\dd s^2_{\text{E}}=f^p\dd t^2 - f^{-p}\dd r^2-r^2f^{1-p}\dd \Omega^2,\\
&\frac{\sigma}{\sqrt{12\phi_0}}=-\frac{1\mp f^{1/k_+}}{1\pm f^{1/k_+}},\qquad \phi =\pm \frac{4\phi_0f^{1/k_+}}{(1\pm f^{1/k_+})^2},\nn \\
&p\equiv \frac{\eta}{k_+}= \frac{\eta}{\sqrt{\eta^2+3}}<1,\nn
\end{align}
where the upper sign corresponds to the solution obtained in the harmonic gauge~\eqref{9} and the lower sign corresponds to the solution obtained by the conformal transformation~\eqref{6c}.

Now, since $p<1$ the solution in the Einstein frame~\eqref{metric2} is not analytic\footnote{For analytic solutions, $p$ should be a positive integer~\cite{cold1,cold2}.} and it cannot be extended for $r<2k_+$. Since the area of the surface $r=2k_+$ (proportional to $r^2f^{1-p}|_{r=2k_+}$) vanishes, $r=2k_+$ is a point-like naked singularity. This can be seen from the expression of the Kretchmann scalar, which diverges in the limit $r\to (2k_+)^+$ as
\begin{equation}
R_{\mu\nu\alpha\beta}R^{\mu\nu\alpha\beta}\propto f^{-2(2-p)}\to\infty .
\end{equation}
The invariant scalar also diverges: $R\propto f^{-(2-p)}\to\infty$.

For the case $\eta =\infty$ ($p=1$), the solution~\eqref{metric2} is the Schwarzschild black hole. The corresponding upper-sign solution in the Jordan frame is defined and the lower-sign solution is undefined.

It is worth mentioning that the parameters ($\eta,\,p$) of the solutions~\eqref{metric2} are still unrestricted real numbers while the corresponding parameters of the solutions derived in~\cite{duality,conf} have some restrictions\footnote{The solution derived in Eq.~(27) of Ref.~\cite{conf} should have the coefficient $(w^{\beta}\pm w^{-\beta})^2$.}. The solutions~\eqref{metric2} constitute thus extensions of  previously derived solutions.\\

\subsection{\textbf{Dynamical normal solutions: $\pmb{C_3=\ell\pi+\lambda\,\ii}$, $\pmb{\ell\in\mathbb{Z}}$, $\pmb{\lambda\in\mathbb{R}}$.}} For this value of $C_3$ the final expressions of ($y,\,\e^{2\beta},\,\e^{2\alpha}=\e^{4\beta}\e^{2\gamma}$) are
\begin{align}
&\label{18a}y=x -\eta -k_+\tanh \big(k_+\,\text{arctanh}\,x\big),\\
&\label{18b}\e^{2\beta}=-\frac{V_0}{12\phi^2}~k_+^2\text{sech}^2\big(k_+\,\text{arctanh}\,x\big)~\e^{-2\gamma},\\
&\label{18c}\e^{2\alpha}=\frac{V_0^2}{12^2\phi^4}~k_+^4\text{sech}^4\big(k_+\,\text{arctanh}\,x\big)~\e^{-2\gamma}.
\end{align}
Since the equations~\eqref{6} are invariant under the change $g_{\mu\nu}\to -g_{\mu\nu}$, we multiply the metric coefficients by $-1$ and change $u\to t$ and $t\to u$, we obtain the dynamical normal metric
\begin{align}\label{8b}
&\dd s^2_{\text{J}}=\e ^{2\alpha(t)}\dd t^2-\e ^{2\gamma(t)}\dd u^2 - (-\e ^{2\beta(t)})\dd \Omega^2,\nn\\
&\sigma =t,\qquad \phi =V_0-\frac{\sigma^2}{12}=V_0-\frac{t^2}{12}
\end{align}
where $t$ is timelike and the other coordinates are spacelike. Here $\e ^{2\gamma(t)}$ is given by~\eqref{16a} where $\sigma$ is to be replaced by $t$ (recall $C=1$ and $\sigma=u$) and ($-\e ^{2\beta(t)},\,\e ^{2\alpha(t)}$) are given by~(\ref{18b}, \ref{18c}) with $\sigma$ replaced by $t$ as follows
\begin{align}\label{8c}
&\e ^{2\gamma(t)}=\frac{1}{12 V_0-t^2 } \Big(\frac{\sqrt{12 V_0}+t}{\sqrt{12 V_0}-t}\Big)^{\eta },\nn\\
&-\e^{2\beta(t)}=\frac{12V_0}{(12 V_0-t^2)^2}~k_+^2\text{sech}^2\Big[k_+\,\text{arctanh}\Big(\frac{t }{\sqrt{12 V_0}}\Big)\Big]\nn\\
& \qquad \quad  \times \e^{-2\gamma(t)},\\
&\e^{2\alpha(t)}=\frac{12^2V_0^2}{(12 V_0-t^2)^4}~k_+^4\text{sech}^4\Big[k_+\,\text{arctanh}\Big(\frac{t }{\sqrt{12 V_0}}\Big)\Big]\nn\\
& \qquad \quad  \times \e^{-2\gamma(t)}.\nn
\end{align}

The other dynamical normal solution obtained by the conformal transformation~\eqref{6c} is
\begin{align}\label{8d}
&\dd s^2_{\text{J}}=\frac{t^2}{12V_0}\Big(\e ^{2\alpha(t)}\dd t^2-\e ^{2\gamma(t)}\dd u^2 - (-\e ^{2\beta(t)})\dd \Omega^2\Big),\nn\\
&\sigma =\frac{12V_0}{t},\qquad \phi =V_0-\frac{\sigma^2}{12}=V_0-\frac{12V_0^2}{t^2},
\end{align}
where ($\e ^{2\gamma(t)},\,-\e ^{2\beta(t)},\,\e ^{2\alpha(t)}$) are given by~\eqref{8c}.

On defining the time $\tau$ by
\begin{equation}
\e^{-2\tau} = \frac{\sqrt{12 V_0}+t}{\sqrt{12 V_0}-t},
\end{equation}
we bring the metrics~(\ref{8b}, \ref{8d}) to the following forms, respectively, with the same metric solution in the Einstein frame
\begin{align}\label{8e}
&\dd s^2_{\text{J}}=\cosh^2\tau~\dd s^2_{\text{E}},\;\phi =\frac{V_0}{\cosh^2\tau},\; \frac{\sigma}{\sqrt{12V_0}}=-\tanh \tau ,\nonumber\\
&\dd s^2_{\text{J}}=\sinh^2\tau~\dd s^2_{\text{E}},\;\phi =\frac{V_0}{\sinh^2\tau},\; \frac{\sigma}{\sqrt{12V_0}}=-\coth \tau ,\nonumber\\
&\dd s^2_{\text{E}}=\frac{k_+^4\e ^{2\eta\tau}}{\cosh^4(k_+\tau)}~\dd \tau^2 - \e ^{-2\eta\tau}\dd u^2 - \frac{k_+^2\e ^{2\eta\tau}}{\cosh^2(k_+\tau)}~\dd \Omega^2,
\end{align}
where we have made the substitution $u\to \sqrt{12V_0}\,u$.

\section{Phantom soluions}
We assume that~\eqref{15b} holds so that $\phi_0 = - V_0$ and the function $\phi$ reads
\begin{equation}\label{sub}
\phi =-V_0 -\frac{\sigma^2}{12}<0.
\end{equation}
Eq.~(\ref{11} is solved by
\begin{equation}\label{p1}
\e^{2 \gamma }=-\frac{1}{12 \phi } \e^{-2\eta \arctan x},
\end{equation}
where $\eta$ is as defined in~\eqref{def}. The case $3-\eta^2>0$ yields static solutions and the case $3-\eta^2<0$ yields dynamical solutions. We exclude the case $3-\eta^2=0$ from our analysis for it leads to an undefined metric with $\e^{2 \beta}\equiv 0$ ($z=\sigma -C_2$ is linear in this case).

\subsection{\textbf{Static phanton solutions: $\pmb{3-\eta^2>0}$.}} In this case the final expressions of ($y,\,\e^{2\beta},\,\e^{2\alpha}=\e^{4\beta}\e^{2\gamma}$) after removing unnecessary constants are
\begin{align}
&\label{p2}y=x -\eta-q \tanh \big(q\,\text{arctan}\,x\big),\\
&\label{p3}\e^{2\beta}=\frac{V_0}{12\phi^2}~q^2\text{sec}^2\big(q\,\text{arctan}\,x\big)~\e^{-2\gamma},\\
&\label{p4}\e^{2\alpha}=\frac{\phi_0^2}{12^2\phi^4}~q^4\text{sec}^4\big(q\,\text{arctan}\,x\big)~\e^{-2\gamma},
\end{align}
where
\begin{equation}\label{q}
q\equiv \sqrt{3-\eta^2}.
\end{equation}
In terms of the new radial coordinate,
\begin{equation}\label{rho}
\rho =\arctan x,
\end{equation}
the metric takes the form
\begin{align}\label{p5}
&\dd s^2_{\text{J}}=\cos^2\rho~\dd s^2_{\text{E}},\;\phi =-\frac{V_0}{\cos^2\rho},\; x=\tan \rho ,\nonumber\\
&\dd s^2_{\text{E}}=\e ^{-2\eta\rho}\dd t^2 - \frac{q^2\e ^{2\eta\rho}}{\cos^2(q\rho)}\Big(\frac{q^2\dd \rho^2}{\cos^2(q\rho)} +\dd \Omega^2\Big),
\end{align}
where we have made the substitution $t\to \sqrt{12V_0}\,t$. The metric obtained by the conformal transformation~\eqref{6c} takes also a simple form with the same metric solution in the Einstein frame
\begin{align}\label{p5b}
&\dd s^2_{\text{J}}=\sin^2\rho~\dd s^2_{\text{E}},\;\phi =-\frac{V_0}{\sin^2\rho},\; x=\cot \rho ,
\end{align}
where $\dd s^2_{\text{E}}$ is the same as in~\eqref{p5}.

We bring~\eqref{p5} to the form derived in~\cite{stt} by the coordinate transformation
\begin{equation}\label{c1}
\bar{q}\bar{\rho}=\frac{\pi}{2}+q\rho,
\end{equation}
yielding
\begin{equation}\label{c2}
\dd s^2_{\text{E}}=\e ^{-2\bar{\eta}\bar{\rho}}\dd t^2 - \frac{\bar{q}^2\e ^{2\bar{\eta}\bar{\rho}}}{\sin^2(\bar{q}\bar{\rho})}\Big(\frac{\bar{q}^2\dd \bar{\rho}^2}{\sin^2(\bar{q}\bar{\rho})} +\dd \Omega^2\Big),
\end{equation}
where we have made the substitution $\e ^{\frac{\pi\eta}{2q}}~t\to t$. The new constants ($\bar{q},\,\bar{\eta}$) are defined by
\begin{equation}\label{c3}
\bar{q}^2\equiv q^2 \exp{\Big(\frac{-\pi\eta}{q}\Big)},\qquad\bar{\eta} \equiv \frac{\bar{q}}{q}~\eta .
\end{equation}
Since $-\pi/2<\rho <\pi/2$~\eqref{rho}, we have
\begin{equation}\label{c4}
(1-q)~\frac{\pi}{2}<\bar{q}\bar{\rho} <(q+1)~\frac{\pi}{2},
\end{equation}
and this makes it clear that the interval of $q$, $0<1\leq \sqrt{3}$~\eqref{q}, needs be divided into two sub-intervals and the case $q=1$.

\subparagraph*{\textbf{A1. $\pmb{0<q<1}$.}}
This case was not discussed in~\cite{stt}. For this sub-interval $\sin(\bar{q}\bar{\rho})$ is never 0 and the metric non-singular. The solution has no spatial infinity since the spherical radius $r\equiv\bar{q}\e ^{\bar{\eta}\bar{\rho}}/\sin(\bar{q}\bar{\rho})$ does not diverge at the end points of $\bar{\rho}$. Similar solutions with no asymptotic flatness occur in the Pleba\'{n}ski--Demia\'{n}ski family of black hole-like and other spacetimes~\cite{no}. The spherical radius $r$ has a minimum vale at $\cot(\bar{q}\bar{\rho}_{\text{min}})=\bar{\eta}/\bar{q}=\eta/q$ yielding
\begin{equation}\label{c5}
\bar{\rho}_{\text{min}}=\text{arccot}\Big[\frac{\eta}{q^2} \exp{\Big(\frac{\pi\eta}{2q}\Big)}\Big],
\end{equation}
where the range of $\text{arccot}$ in the open interval ($0,\pi$). The radius of the throat of the wormhole is
\begin{equation}\label{th}
r_{\text{th}}=\sqrt{3}~\exp{\Big\{\frac{\eta }{q}\Big[\text{arccot}\Big(\frac{\eta }{q}\Big)-\pi\Big]\Big\}}.
\end{equation}
Thus the metric~\eqref{c2} is a non-asymptotically flat wormhole.

\subparagraph*{\textbf{A2. $\pmb{q=1}$.}} The metric has now two spatial infinities at $\bar{q}\bar{\rho}=0$ and $\bar{q}\bar{\rho}=\pi$ where the spherical radius $r$ diverges. The latter has the same minimum value at $\bar{\rho}_{\text{min}}$ given by~\eqref{th}.

As we shall see the two asymptotic regions have different masses. The mass $M$ of the wormhole is defined by\footnote{The parameters of Ref.~\cite{stt} and those used in this work are related by: $m^2=\eta^2\e ^{-2\pi\eta /q}$, $k^2=q^2\e ^{-2\pi\eta /q}$, $C^2=6\e ^{-2\pi\eta /q}$ up to a proportionality factor.}
\begin{equation}\label{c6}
\frac{\bar{q}^2\e ^{2\bar{\eta}\bar{\rho}}}{\sin^2(\bar{q}\bar{\rho})}\Big(\frac{\bar{q}^2\dd \bar{\rho}^2}{\sin^2(\bar{q}\bar{\rho})} +\dd \Omega^2\Big)\underset{(r\to \infty)}{\simeq}\frac{\dd r^2}{1-\frac{2M}{r}}+r^2\dd \Omega^2.
\end{equation}
In the limit $\bar{q}\bar{\rho}\to 0^+$ ($r\simeq 1/\bar{\rho}\to +\infty$) we find that $M=\bar{\eta}$ and in the limit  $\bar{q}\bar{\rho}\to \pi^-$ [$r\simeq \bar{q}\e ^{\pi\eta/q}/(\bar{q}\bar{\rho}-\pi)\to -\infty$] we find a negative mass $M=-\bar{\eta}\e ^{\pi\eta/q}$. Recall that we are restricting the sign of $\eta \geq 0$~\eqref{ass}; had we not done that, that is, whatever the sign of $\eta$ were, the two masses would always have opposite signs. The wormhole is attractive if $M >0$ and repulsive if $M <0$.

\subparagraph*{\textbf{A3. $\pmb{1< q\leq \sqrt{3}}$.}}
To each value of $q$ in this range correspond three solutions depending on how we restrict the domain of definition of $\bar{\rho}$.

If $(1-q)\pi/2<\bar{q}\bar{\rho} < 0$, the solution has only one spatial infinity ($r\to \infty$) at $\bar{q}\bar{\rho} = 0$ and the spherical radius $r$ is bounded from below by $q \e ^{-\pi\eta(1+q)/(2q)}/\cos(q\pi /2)$. This solution represents thus the exterior geometry of a star.

If $0<\bar{q}\bar{\rho} < \pi$, the solution is a wormhole with two asymptotic regions at $\bar{q}\bar{\rho}=0$ and $\bar{q}\bar{\rho}=\pi$ as described in the case A2.

If $\pi<\bar{q}\bar{\rho} < (q+1)\pi/2$, the solution has only one spatial infinity ($r\to \infty$) at $\bar{q}\bar{\rho} = \pi$ and the spherical radius $r$ is bounded from below by $q \e ^{-\pi\eta(1-q)/(2q)}/\cos(q\pi /2)$. This solution represents also the exterior geometry of a star.

\subsection{\textbf{Dynamical phanton solutions: $\pmb{3-\eta^2<0}$.}} Let
\begin{equation}\label{qb}
k_-\equiv \sqrt{\eta^2-3}.
\end{equation}
The expressions of ($y,\,\e^{2\beta},\,\e^{2\alpha}=\e^{4\beta}\e^{2\gamma}$) are
\begin{align}
&\label{19a}y=x -\eta +k_-\tanh \big(k_-\,\arctan x\big),\\
&\label{19b}\e^{2\beta}=-\frac{V_0}{12\phi^2}~k_-^2\text{sech}^2\big(k_-\,\arctan x\big)~\e^{-2\gamma},\\
&\label{19c}\e^{2\alpha}=\frac{V_0^2}{12^2\phi^4}~k_-^4\text{sech}^4\big(k_-\,\arctan x\big)~\e^{-2\gamma}.
\end{align}
To obtain the dynamical solution we multiply the above metric coefficients by $-1$ and change $u\to t$ and $t\to u$, we obtain the dynamical metric
\begin{align}\label{9b}
&\dd s^2_{\text{J}}=\e ^{2\alpha(t)}\dd t^2-\e ^{2\gamma(t)}\dd u^2 - (-\e ^{2\beta(t)})\dd \Omega^2,\nn\\
&\sigma =t,\qquad \phi =-V_0-\frac{\sigma^2}{12}=-V_0-\frac{t^2}{12}
\end{align}
where $t$ is timelike and the other coordinates are spacelike. Here $\e ^{2\gamma(t)}$ is given by~\eqref{p1} where $\sigma$ is to be replaced by $t$ (recall $C=1$ and $\sigma=u$) and ($-\e ^{2\beta(t)},\,\e ^{2\alpha(t)}$) are given by~(\ref{19b}, \ref{19c}) with $\sigma$ replaced by $t$ as follows
\begin{align}\label{9c}
&\e ^{2\gamma(t)}=\frac{1}{12 V_0+t^2}~\e^{-2\eta \arctan (t/\sqrt{12V_0})},\nn\\
&-\e^{2\beta(t)}=\frac{12V_0}{(12 V_0+t^2)^2}~k_-^2\text{sech}^2\Big[k_-\,\arctan\Big(\frac{t }{\sqrt{12 V_0}}\Big)\Big]\nn\\
& \qquad \quad  \times \e^{-2\gamma(t)},\\
&\e^{2\alpha(t)}=\frac{12^2V_0^2}{(12 V_0+t^2)^4}~k_-^4\text{sech}^4\Big[k_-\,\arctan\Big(\frac{t }{\sqrt{12 V_0}}\Big)\Big]\nn\\
& \qquad \quad  \times \e^{-2\gamma(t)}.\nn
\end{align}

The dynamical solution obtained by the conformal transformation~\eqref{6c} is
\begin{align}\label{9d}
&\dd s^2_{\text{J}}=\frac{t^2}{12V_0}\Big(\e ^{2\alpha(t)}\dd t^2-\e ^{2\gamma(t)}\dd u^2 - (-\e ^{2\beta(t)})\dd \Omega^2\Big),\nn\\
&\sigma =\frac{12V_0}{t},\qquad \phi =-V_0-\frac{\sigma^2}{12}=-V_0-\frac{12V_0^2}{t^2},
\end{align}
where ($\e ^{2\gamma(t)},\,-\e ^{2\beta(t)},\,\e ^{2\alpha(t)}$) are given by~\eqref{9c}.

Now, we introduce the time $\tau$ by
\begin{equation}
\tau =\arctan(t/\sqrt{12V_0}).
\end{equation}
The two dynamical phantom solutions take the forms
\begin{align}\label{p6}
&\dd s^2_{\text{J}}=\cos^2\tau~\dd s^2_{\text{E}},\;\phi =-\frac{V_0}{\cos^2\tau},\; \frac{t}{\sqrt{12V_0}}=\tan \tau ,\nonumber\\
&\dd s^2_{\text{E}}=\frac{k_-^4\e ^{2\eta\tau}}{\cosh^4(k_-\tau)}~\dd \tau^2 - \e ^{-2\eta\tau}\dd u^2 - \frac{k_-^2\e ^{2\eta\tau}}{\cosh^2(k_-\tau)}~\dd \Omega^2,
\end{align}
and
\begin{align}\label{p7}
&\dd s^2_{\text{J}}=\sin^2\tau~\dd s^2_{\text{E}},\;\phi =-\frac{V_0}{\sin^2\tau},\; \frac{t}{\sqrt{12V_0}}=\cot \tau ,
\end{align}
where $\dd s^2_{\text{E}}$ is the same as in~\eqref{p6}.

\section{Properties of the dynamical solutions}
In the Einstein frame we have obtained one dynamical normal solution, given by~\eqref{8e}, and one dynamical phantom solution, given by~\eqref{p6}. These are combined into one expression
\begin{equation}\label{pm}
\dd s^2_{\text{E}}=\frac{k_\pm^4\e ^{2\eta\tau}}{\cosh^4(k_\pm\tau)}~\dd \tau^2 - \e ^{-2\eta\tau}\dd u^2 - \frac{k_\pm^2\e ^{2\eta\tau}}{\cosh^2(k_\pm\tau)}~\dd \Omega^2,
\end{equation}
where, from now on, the upper (lower) sign corresponds to normal (phantom) solution. We checked explicitly that these dynamical metrics, with their corresponding fields given in~\eqref{8e} and~\eqref{p6}, are exact solutions to the field equations~\eqref{5} and~\eqref{6}.

It is clear that these two one-parameter solutions are anisotropic for the metric components are not proportional. 

As $\tau\to -\infty$, there is a cosmological singularity where $g_{uu}\to\infty$, $g_{\theta\theta}\to 0$ and $g_{\phi\phi}\to 0$. This could be defined as a {\textquotedblleft radial cigar cosmological singularity\textquotedblright}, since as one looks into any radial direction one sees a radial expansion and a transverse collapse. The cigar as well as other terminologies have been introduced in~\cite{dsc}. The structure of this anisotropy past singularity ($\tau\to -\infty$) is revealed upon introducing the cosmic time parameter $T=\e ^{\tau}\to 0$ as $\tau\to -\infty$ by which the metric~\eqref{pm} approaches the Kasner-like solution~\cite{Kasner}
\begin{multline}\label{pm1}
\dd s^2_{\text{E}}\simeq 16k_\pm^4 T^{2a_1}~\dd T^2 - T^{2a_2}\dd u^2 \\- 4k_\pm^2 (T^{2a_3}~\dd \theta^2+T^{2a_4}\sin^2\theta ~\dd \varphi^2),
\end{multline}
where we have used the expression of $\dd \Omega^2=\dd \theta^2+\sin^2\theta ~\dd \varphi^2$. The parameters $a_i$ ($i=1\to 4$) defined by
\begin{equation}\label{pm3}
a_1\equiv 2k_\pm +\eta -1,\quad a_2\equiv -\eta,\quad a_3=a_4\equiv  k_\pm +\eta\,,
\end{equation}
satisfy the relations
\begin{align}
\label{r1}&a_2+a_3+a_4=1+a_1,\\
\label{r2}&a_2^2+a_3^2+a_4^2=(1+a_1)^2\mp 6>0,
\end{align}
where the right-hand side of~\eqref{r2} is always positive for all $\eta$ (positive or negative). Note that the first relation~\eqref{r1} is identical to Kasner relation~\cite{Kasner}.

There is another anisotropy cosmological future singularity as $\tau\to \infty$ where $g_{uu}\to 0$, $g_{\theta\theta}\to 0$, $g_{\phi\phi}\to 0$ for the normal solution, corresponding to a point singularity~\cite{dsc}, and $g_{uu}\to 0$, $g_{\theta\theta}\to \infty$,  $g_{\phi\phi}\to \infty$ for the phantom solution, corresponding to a {\textquotedblleft radial pancake cosmological singularity\textquotedblright}. To reveal its structure we introduce the cosmic time parameter $T=\e ^{-\tau}\to 0$ as $\tau\to \infty$ by which the metric~\eqref{pm} approaches the Kasner-like solution~\eqref{pm1} where this time the parameters $a_i$ ($i=1\to 4$) are defined by
\begin{equation}\label{pm5}
a_1\equiv 2k_\pm -\eta -1,\quad a_2\equiv \eta,\quad a_3=a_4\equiv  k_\pm -\eta \,,
\end{equation}
and they satisfy the same relations~\eqref{r1} and~\eqref{r2}.

We can also introduce the time $\bar{T}=[4k_\pm^2/(1+a_1)] T^{1+a_1}$, for both past and future singularities, bringing~\eqref{pm1} to
\begin{multline}\label{pm2}
\dd s^2_{\text{E}}\simeq \dd \bar{T}^2 -\Big(\frac{1+a_1}{4k_\pm^2}\Big)^{2p_1} \bar{T}^{2p_1}\dd u^2 \\- 4k_\pm^2 \Big(\frac{1+a_1}{4k_\pm^2}\Big)^{2p_2} (\bar{T}^{2p_2}~\dd \theta^2+\bar{T}^{2p_3}\sin^2\theta ~\dd \varphi^2),
\end{multline}
where $p_1=a_2/(1+a_1)$ and $p_2=p_3=a_3/(1+a_1)$ and
\begin{align}
\label{r3}&p_1+p_2+p_3=1,\\
\label{r4}&p_1^2+p_2^2+p_3^2=1\mp \frac{6}{(1+a_1)^2}>0.
\end{align}

As times runs from $\tau=-\infty$ to $\tau=\infty$ no other cosmological singularity occurs during the anisotropic cosmic evolution. From the point of view of dynamical system analysis, this means that these two singularities, at $\tau=-\infty$ and at $\tau=\infty$, are (generally saddle) equilibrium points behaving as source and sink during the evolution, and the dynamical solution~\eqref{pm} describes a heteroclinic orbit of evolution coinciding with a part of the skeleton of the corresponding phase portrait. An intermediate evolution~\cite{dsc} between these two singularities, not coinciding with the heteroclinic orbit, may take place too where one singularity behaves as a past asymptote (a repeller) and the other as a future asymptote (an attractor).

As is clear from~\eqref{pm2} the expansion or the collapse in the radial direction is generally different from that in the transverse direction as one approaches the singularities. However, isotropic evolution may take place for special values of $\eta$. For the future singularity ($\tau\to \infty$) we observe isotropy only for normal solutions if $p_1=p_2=p_3=1/3$ ($\Rightarrow a_2=a_3$), which results in $\eta=1$ upon using~\eqref{pm5}. For normal solutions a purely transverse expansion or collapse (no radial expansion or collapse) is also possible in the case $\eta = 0$ yielding $p_1=0$ and $p_2=p_3=1/2$ for both singularities. Since it is not possible to have $p_2=p_3=0$, no purely radial expansion can exist.

Since the right-hand side of~\eqref{r4} is always positive for all $\eta$ (positive or negative), we must have $p_1+p_2+p_3=1$ and $p_1^2+p_2^2+p_3^2<1$ for normal solutions, which are known as Jacobs stiff perfect fluid conditions~\cite{dsc,Jacobs}. For phantom solutions, however, we have $p_1+p_2+p_3=1$ and $p_1^2+p_2^2+p_3^2>1$. Both normal and phantom solutions approach Kasner solution $p_1+p_2+p_3=1$ and $p_1^2+p_2^2+p_3^2\to 1^{\mp}$ in the limit $\eta\to\infty$.




%



\end{document}